\documentclass[aps,pre,preprint,groupedaddress,showpacs]{revtex4}
\usepackage{graphicx}
\usepackage{amsmath}
\usepackage{amsfonts}
\begin{document}
\title{Photoionization effects in streamer discharges}
\author{Manuel Array\'as$^{1}$, Marco A. Fontelos$^{2}$ and Jos\'e L. Trueba$^{1}$}
\affiliation{$^{1}$Departamento de F\'{\i}sica Aplicada,
Universidad Rey Juan Carlos, Tulip\'{a}n s/n, 28933 M\'{o}stoles,
Madrid, Spain} \affiliation{$^{2}$Instituto de Matem\'{a}ticas y
F\'{i}sica Fundamental,
  Consejo Superior de Investigaciones Cient\'{i}ficas, C/ Serrano
  123, 28006 Madrid, Spain}
\begin{abstract}
In this paper we study the effects of photoionization processes on
the propagation of both negative and positive streamer discharges.
We show that negative fronts accelerate in the presence of
photoionization events. The appearance and propagation of positive
streamers travelling with constant velocity is explained as the
result of the combined effects of photoionization and electron
diffusion. The photoionization range plays an important role for
the selection of the velocity of the streamer as we show in this
work.
\end{abstract}
\date{\today}
 \maketitle
\section{Introduction}
Since Raether \cite{Raether} used cloud chamber photographs to study
the creation and propagation of streamer discharges there has been
considerable effort to understand the underlying processes driving them. A
streamer discharge is considered to be a plasma channel which
propagates in a gas. The discharge propagates by ionizing the medium
in front of its charged head due to a strong field induced by the head
itself. This kind of discharges produces sharp ionization waves that
propagate into a non-ionized gas, leaving a non-equilibrium plasma
behind.

Raether himself realized that Townsend's mechanism which takes into
account the creation of extra charge by impact ionization \cite{Loeb}
was not enough to explain the velocity of propagation of a streamer
discharge. He pointed to photoionization as the process which
enhances the propagation of the streamer. Due to the recombination of
positive ions and electrons, the head of the discharge is a strong
source of high energetic photons. Photons, emitted by the atoms that
previous collisions have excited, initiate secondary avalanches in the
vicinity of the head which move driven by the local electric field
increasing the velocity of propagation of the front.

In this paper we study the role played by photoionization in the
propagation of both negative and positive streamers. We take a model
widely used in numerical simulations and find an effective simplified
model. We discuss how this simplified model retains all
the physics of streamer discharges including photoionization. The
photoionization is modelled as a nonlocal source term. We take the
case of air and consider optical emissions from N$_2$ and N$_2^+$
molecules. Then we consider the sole role of photoionization in
negative planar shock fronts. Finally we analyse the case of positive
planar fronts and propose a mechanism for their formation and
propagation. We end with an analysis of results and conclusions.

\section{Model for a streamer discharge}
Here we consider a fluid description of a low-ionized plasma based
on kinetic theory. The balance equation for the particle density
of electrons $N_{e}$ is the lowest moment of the Boltzmann
equation,
\begin{equation}
\frac{\partial N_{e}}{\partial \tau} + \nabla_{\bf R} \cdot \left( N_{e}
{\bf U}_{e} \right) = S_{e},
\label{balance1}
\end{equation}
where ${\bf R}$ is the position vector, $\tau$ is time,
$\nabla_{\bf R}$ is the gradient in configuration space, ${\bf
U}_{e} ({\bf R}, \tau)$ is the average (fluid) velocity of
electrons and $S_{e}$ is the source term, i.e. the net creation
rate of electrons per unit volume as a result of collisions. It is
convenient to define the electron current density ${\bf J}_{e}
({\bf R}, \tau)$ as
\begin{equation}
{\bf J}_{e} ({\bf R}, \tau) = N_{e} ({\bf R}, \tau)\, {\bf U}_{e}
({\bf R}, \tau), \label{def-curr}
\end{equation}
so that the balance equation can also be written as
\begin{equation}
\frac{\partial N_{e}}{\partial \tau} + \nabla_{\bf R} \cdot {\bf
J}_{e} = S_{e}. \label{balance-electrons}
\end{equation}
The same procedure can be done, in principle, for positive
($N_{p}$) and negative ($N_{n}$) ion densities to give
\begin{eqnarray}
\frac{\partial N_{p}}{\partial \tau} + \nabla_{\bf R} \cdot {\bf
J}_{p} = S_{p}, \label{balance-positive} \\
\frac{\partial N_{n}}{\partial \tau} + \nabla_{\bf R} \cdot {\bf
J}_{n} = S_{n}, \label{balance-negative}
\end{eqnarray}
where ${\bf J}_{p,n}$ are the current densities of positive and
negative ions, respectively, and $S_{p,n}$ are source terms.
Conservation of charge has to be imposed in all processes, so that
the condition
\begin{equation}
S_{p} = S_{e} + S_{n}, \label{cons-charge}
\end{equation}
holds for the source terms. Some physical approximations can now
be done in order to simplify the balance equations
(\ref{balance-electrons})--(\ref{balance-negative}). The first
one is to assume that the electron current ${\bf J}_e$ is
approximated as the sum of a drift (electric force) and a
diffusion term
\begin{equation}
{\bf J}_e=-\mu_e {\boldsymbol{\cal E}} N_e - D_e \nabla_{\bf R}
N_e , \label{el-current}
\end{equation}
where ${\boldsymbol{\cal E}}$ is the total electric field (the sum
of the external electric field applied to initiate the propagation
of a ionization wave and the electric field created by the local
point charges) and $\mu_e$ and $D_e$ are the mobility and
diffusion coefficient of the electrons. Note that, as the initial
charge density is low and there is no applied magnetic field, the
magnetic effects in equation (\ref{el-current}) are neglected.
Concerning the diffusion coefficient, in the case of equilibrium,
the kinetic theory of gases links diffusion to mobility through
Einstein's relation $D_e/\mu_e = kT/e$. With respect to positive
and negative ions, on time-scales of interest for the case of
 streamer discharges, the ion currents can be neglected
because they are more than two orders of magnitude smaller than
the electron ones \cite{AJP}, so we will take
\begin{equation}
{\bf J}_{p}= {\bf J}_{n} =0. \label{ion-current}
\end{equation}
Consider now the processes that give rise to the source terms
$S_{e,p,n}$:
\begin{enumerate}
\item  The first of these processes is the creation of free electrons
by impact ionization: an electron is accelerated in a strong local
field, collides with a neutral molecule and ionizes it. The result
is the generation of new free electrons and a positive ion. The
ionization rate is given by
\begin{equation}
S_{e}^{i} = S_{p}^{i} =  \nu_{i} N_e , \label{impact}
\end{equation}
where the ion production rate $\nu_{i}$ depends on the local
electric field, the density of the neutral particles of the gas
and their effective ionization cross sections.
\item  The second possible process is attachment: when an electron
collides with a neutral gas atom or molecule, it may become
attached, forming a negative ion. This process depends on the
energy of the electron and the nature of the gas \cite{Dhali}. The
attachment rate can be written as
\begin{equation}
S_{n}^{a} = - S_{e}^{a} =  \nu_{a} N_e , \label{attachment}
\end{equation}
where $\nu_{a}$ is the attachment rate coefficient. Note that the
creation of negative ions due to these processes reduces the
number of free electrons, so $S_{e}^{a}$ is negative.
\item  There are also two possible kinds of recombination processes:
a free electron with a positive ion and a negative
ion with a positive ion. The recombination rate is
\begin{equation}
S_{e}^{ep} =  S_{p}^{ep} =  - \nu_{ep} N_{e} N_{p} , \label{recombination1}
\end{equation}
for electron-positive ion recombination, and
\begin{equation}
S_{p}^{np} =  S_{n}^{np} =  - \nu_{np} N_{n} N_{p} , \label{recombination2}
\end{equation}
for positive ion-negative ion recombination, $\nu_{ep}$ and
$\nu_{np}$ being the recombination coefficients respectively.
\item  Finally,  we can include photoionization: photons created
by recombination or scattering processes can interact with a
neutral atom or molecule, producing a free electron and a
positive ion. Models for the creation rate of electron-positive
ion pairs due to photoionization are non-local. This rate will be
here denoted by
\begin{equation}
S_{e}^{ph} = S_{p}^{ph} = S^{ph} . \label{photoionization}
\end{equation}
\end{enumerate}
Taking into account the expressions (\ref{el-current}) and
(\ref{ion-current}) for the current densities, and the equations
(\ref{impact})--(\ref{photoionization}) for the source terms, we
obtain a deterministic model for the evolution of the streamer
discharge,
\begin{eqnarray}
\frac{\partial N_{e}}{\partial \tau} &=& \nabla_{\bf R} \cdot
\left( \mu_e {\boldsymbol{\cal E}} N_e + D_e \nabla_{\bf R} N_e
\right) + \nu_{i} N_e \nonumber \\&-& \nu_{a} N_e  - \nu_{ep}
N_{e} N_{p} + S^{ph} ,
\label{model1} \\
\frac{\partial N_{p}}{\partial \tau} &=& \nu_{i} N_e - \nu_{ep} N_{e} N_{p}
- \nu_{np} N_{n} N_{p} + S^{ph} , \label{model2} \\
\frac{\partial N_{n}}{\partial \tau} &=& \nu_{a} N_e - \nu_{np} N_{n} N_{p} . \label{model3}
\end{eqnarray}
In order for the model to be complete, it is necessary to give
expressions for the source coefficients $\nu$, the electron
mobility $\mu_{e}$, the diffusion coefficient $D_{e}$ and the
photoionization source term $S^{ph}$. Finally, we have to impose
equations for the evolution of the electric field
${\boldsymbol{\cal E}}$. This evolution of the electric field is
given by Poisson's equation,
\begin{equation}
\nabla_{\bf R}\cdot{\boldsymbol{\cal E}} =
\frac{e}{\varepsilon_{0}} \, \left( N_{p} - N_{n} -N_{e} \right) ,
\label{poisson}
\end{equation}
where $e$ is the absolute value of the electron charge,
$\varepsilon_{0}$ is the permittivity of the gas, and we are
assuming that the absolute value of the charge of positive and
negative ions is $e$. Note that the coupling between the space
charges and the electric field in the model makes the problem
nonlinear. The model  given by (\ref{model1}), (\ref{model2}), and
(\ref{model3}),  together with (\ref{poisson}) has been studied
numerically in the literature \cite{liu}. There are other works
where the electrical current due to ions (\ref{ion-current}) is
taken into account although not photoionization \cite{Vit}.

\section{A simplified model}
In this section we will simplify the model given by equations
(\ref{model1})--(\ref{model3}).  In order to be specific and fix ideas
we shall consider the case of air. In \cite{liu}, some data are
presented for the ionization coefficients and the photoionization
source term. Using these data we shall see that one can neglect
the quadratic terms involving the coefficients $\nu_{ep}$ and
$\nu_{np}$ since they are about two orders of magnitude smaller than
$\nu_{i}$. The same can be said about the terms involving the
coefficient $\nu_{a}$. First we write equations
(\ref{model1})--(\ref{model3}) as
\begin{eqnarray}
\frac{\partial N_{e}}{\partial \tau} &=& \nabla_{\bf R} \cdot
\left( \mu_e {\boldsymbol{\cal E}} N_e + D_e \nabla_{\bf R} N_e
\right) \nonumber \\&+& \left( \nu_{i} - \nu_{a} - \nu_{ep} N_{p}
\right) N_{e} + S^{ph} ,
\label{model1ren} \\
\frac{\partial N_{p}}{\partial \tau} &=& \left( \nu_{i} - \nu_{ep}
N_{p} \right) N_{e}
- \nu_{np} N_{n} N_{p} + S^{ph} , \label{model2ren} \\
\frac{\partial N_{n}}{\partial \tau} &=& \nu_{a} N_e - \nu_{np}
N_{n} N_{p} . \label{model3ren}
\end{eqnarray}
In these equations, and using the data in \cite{liu} (Figure 1 and
Table 2), the term $\nu_{i}$ is of the order of $10^{10} \,
\mbox{s}^{-1}$ for large electric fields, $\nu_{a}$ is about
$10^{8} \, \mbox{s}^{-1}$, and $\nu_{ep}$ and $\nu_{np}$ are about
$10^{-13} \, \mbox{m}^{3} \cdot \mbox{s}^{-1}$. Moreover, $N_{p}$
is of the same order of $N_{e}$. Then, in equation
(\ref{model3ren}), in the stationary regime when the particle
densities reach the saturation values, one has $N_{n} \sim
\nu_{a}/\nu_{np} \sim 10^{21} \mbox{m}^{-3}$. So that, it follows
from equation (\ref{model2ren}) that, in the stationary regime,
the term $\nu_{np} N_{n} N_{p} \sim 10^{8} N_{p}$ is two orders of
magnitude smaller than the term $\nu_{i} N_{e} \sim 10^{10}
N_{e}$. Hence the terms $\nu_{a} N_{e}$ and $\nu_{np} N_{n} N_{p}$
can safely be neglected. The model then reads
\begin{eqnarray}
\frac{\partial N_{e}}{\partial \tau} &=& \nabla_{\bf R} \cdot
\left( \mu_e {\boldsymbol{\cal E}} N_e + D_e \nabla_{\bf R} N_e
\right) \nonumber \\&+& \left( \nu_{i} - \nu_{ep} N_{p} \right)
N_{e} + S^{ph} ,
\label{model1ren2} \\
\frac{\partial N_{p}}{\partial \tau} &=& \left( \nu_{i} - \nu_{ep}
N_{p} \right) N_{e} + S^{ph} . \label{model2ren2}
\end{eqnarray}
In order to neglect the term $\nu_{ep} N_{e} N_{p}$ by comparison
with the term $\nu_{i} N_{e}$, it is necessary than $N_{p}$ (and
then $N_{e}$) satisfies $N_{p} \ll \nu_{i}/\nu_{ep} \sim 10^{23}
\mbox{m}^{-3}$. To see that it is the case, we use the Poisson equation
(\ref{poisson}) to write equation (\ref{model1ren2}), without the
term $\nu_{ep} N_{e} N_{p}$, as
\begin{eqnarray}
\frac{\partial N_{e}}{\partial \tau} &-& \mu_{e} {\boldsymbol{\cal
E}} \cdot \nabla_{\bf R} N_e  - D_e \nabla_{\bf R}^{2} N_e
\nonumber \\ &=& \left( \nu_{i} + \mu_{e}
\frac{e}{\varepsilon_{0}} \left( N_{p} - N_{e} \right) \right)
N_{e} +  S^{ph}. \label{ren1}
\end{eqnarray}
From this expression, looking at its RHS, we can see that, while
$S^{ph}$ has small effect and the total populations of both ions
and electrons, $N_{e}$ can grow only up to a saturation value at
which $ \nu_{i} + \mu_{e} \frac{e}{\varepsilon_{0}} \left( N_{p} -
N_{e} \right) = 0$, i.e.
\begin{equation}
N_{e} - N_{p} \leq \frac{\nu_{i}}{\mu_{e} e/\varepsilon_{0}} \sim
10^{20} \mbox{m}^{3}, \label{ren2}
\end{equation}
at all times. Therefore neither $N_{p}$ nor $N_{e}$ reach values
close to $10^{23} \mbox{m}^{-3}$, and all the assumptions which
led to neglect $\nu_{ep} N_{e} N_{p}$ are justified. Our
simplified model will be
\begin{eqnarray}
\frac{\partial N_{e}}{\partial \tau} &=& \nabla_{\bf R} \cdot
\left( \mu_e {\boldsymbol{\cal E}} N_e + D_e \nabla_{\bf R} N_e
\right) + \nu_{i} N_e  + S^{ph} ,
\label{model1sim} \\
\frac{\partial N_{p}}{\partial \tau} &=& \nu_{i} N_e + S^{ph}.
\label{model2sim}
\end{eqnarray}
Let us remark that the orders of magnitude deduced for $N_{e}$ and
$N_{p}$ coincide with those found in full numerical simulations by
Liu and Pasko \cite{liu}.

\section{The photoionization term}
In this section we will write down an explicit form of the
photoionization source term. In our study on the effects of
photoionization on the evolution of streamers in air we
consider that only optical emissions from ${\mbox N}_{2}$ and ${\mbox
  N}_{2}^{+}$ molecules can ionize ${\mbox O}_{2}$ molecules. The
photoionization rate, due to the fact that the number of photons
emitted is physically proportional to the number of ions produced by
impact ionization, is written as the following nonlocal source term
\cite{liu,naidis},
\begin{equation}
S_{ph} ({\bf R})= S_{0} \int \nu_{i} ({\bf R}^{\prime}) N_{e}
({\bf R}^{\prime}) \, K_{ph} (|{\bf R} - {\bf R}^{\prime}|) \, d^3
R^{\prime}, \label{liu1}
\end{equation}
where $S_{0}$ is given by
\begin{equation}
S_{0} = \frac{1}{4 \pi} \frac{p_{q}}{p + p_{q}} \, \xi \left(
\frac{\nu_{*}}{\nu_{i}} \right) \frac{1}{\ln{(\chi_{max} /
\chi_{min})}}. \label{liu1b}
\end{equation}
In this expression, $p_{q}$ is the quenching pressure of the
single states of ${\mbox N}_{2}$, $p$ is the gas pressure, $\xi$
is the average photoionization efficiency in the interval of
radiation frequencies relevant to the problem, $\nu_{*}$ is the
effective excitation coefficient for ${\mbox N}_{2}$ state
transitions from which the ionization radiation comes out (we take
$\nu_{*}/\nu_{i}$ to be a constant), and $\chi_{min}$ and
$\chi_{max}$ are, respectively, the minimum and maximum absorption
cross sections of ${\mbox O}_{2}$ in the relevant radiation
frequency interval. The kernel $K_{ph} (|{\bf R} - {\bf
R}^{\prime}|)$ is written as \cite{Zhe}
\begin{equation}
K_{ph} (R) = \frac{\exp{(-\chi_{1} R)} - \exp{(-\chi_{2}
R)}}{R^{3}}, \label{liu3}
\end{equation}
in which $\chi_{1} = \chi_{min} p_{O_{2}}$ and $\chi_{2} =
\chi_{max} p_{O_{2}}$, so that $\chi_{1} < \chi_{2}$.
For the ionization coefficient $\nu_{i}$, we take the
phenomenological approximation given by Townsend \cite{Loeb},
\begin{equation}
\nu_{i} = \mu_e |{\boldsymbol{\cal E}}| \alpha_0 \exp \left(
\frac{-{\cal E}_0}{|{\boldsymbol{\cal E}}|} \right),
\label{townsend}
\end{equation}
where $\mu_e$ is the electron mobility, $\alpha_{0}$ is the
inverse of ionization length, and ${\cal E}_{0}$ is the
characteristic impact ionization electric field. Note also that
$\mu_e|{\boldsymbol{\cal E}}|$ is the drift velocity of electrons.
Townsend approximation provides some physical scales and intrinsic
parameters of the model. It is then convenient to reduce the
equations to dimensionless form. Natural units are given by the
ionization length $R_0=\alpha_0^{-1}$, the characteristic impact
ionization field ${\cal E}_0$, and the electron mobility $\mu_e$,
which lead to the velocity scale $U_0=\mu_e {\cal E}_0$, and the
time scale $\tau_0=R_0/U_0$. We introduce the dimensionless
variables ${\bf r}={\bf R}/R_0$, $t=\tau/\tau_0$, the
dimensionless field ${\bf E}={\boldsymbol{\cal E}}/{\cal E}_0$,
the dimensionless electron and positive ion particle densities
$n_e=N_e/N_0$ and $n_p=N_p/N_0$ with $N_0=\varepsilon_0 {\cal
E}_0/(e R_0)$, and the dimensionless diffusion constant
$D=D_e/(R_0 U_0)$. The dimensionless model reads then,
\begin{eqnarray}
\frac{\partial n_{e}}{\partial t} &=& \nabla \cdot \left( n_{e}
{\bf E} + D \nabla n_e \right) + n_e |{\bf E}| e^{-1/|{\bf E}|} +
S ,
\label{model1sindim} \\
\frac{\partial n_{p}}{\partial t} &=& n_e |{\bf E}| e^{-1/|{\bf
E}|} + S . \label{model2sindim}
\end{eqnarray}
where $S$ is the dimensionless photoionization source term,
\begin{equation}
S ({\bf r})= S_{0} \int n_e ({\bf r}^{\prime}) |{\bf E} ({\bf
r}^{\prime})| e^{-1/|{\bf E} ({\bf r}^{\prime})|} \, K (|{\bf r} -
{\bf r}^{\prime}|) \, d^3 r^{\prime} , \label{model3sindim}
\end{equation}
and
\begin{equation}
S_{0} = \frac{1}{4 \pi} \frac{p_{q}}{p + p_{q}} \, \xi \left(
\frac{\nu_{*}}{\nu_{i}} \right) \frac{1}{\ln{(\chi_{max} /
\chi_{min})}}. \label{liu1bb}
\end{equation}
Also,
\begin{equation} K (r) = \frac{\exp{(-(\chi_{1}/\alpha_{0}) r)} -
\exp{(-(\chi_{2}/\alpha_{0}) r)}}{r^{3}}. \label{model4sindim}
\end{equation}
In this paper, we restrict ourselves to a planar geometry, in
which the evolution of the ionization front is along the $z$-axis.
In this case, the photoionization source term can be written as
\begin{equation}
S (z)= S_{0} \int dz^{\prime} \, n_e (z^{\prime},t) |{\bf E}
(z^{\prime},t)| e^{-1/|{\bf E} (z^{\prime},t)|} \, I
(|z-z^{\prime}|) , \label{march1}
\end{equation}
where
\begin{eqnarray}
I (|z-z^{\prime}|) &=& \int_{-\infty}^{\infty} dy^{\prime}
\int_{-\infty}^{\infty} dx^{\prime} \frac{1}{(x^{\prime ^2} +
y^{\prime 2} + (z-z^{\prime})^{2})^{3/2}} \nonumber \\ &\times &
\left( e^{(-(\chi_{1}/ \alpha_{0}) \sqrt{x^{\prime 2} + y^{\prime
2} + (z-z^{\prime})^{2}})} - e^{(- (\chi_{2}/ \alpha_{0})
\sqrt{x^{\prime 2} + y^{\prime 2} + (z-z^{\prime})^{2}})} \right).
\label{march2}
\end{eqnarray}
Changing to cylindrical coordinates, and integrating in the polar
angle, equation (\ref{march2}) results in
\begin{equation}
I (|z-z^{\prime}|) = 2 \pi \int_{0}^{\infty} r\,dr\, \frac{\exp{(- (
\chi_{1}/ \alpha_{0}) \sqrt{r^2 + (z-z^{\prime})^{2}})} - \exp{(-
( \chi_{2}/ \alpha_{0}) \sqrt{r^2 + (z-z^{\prime})^{2}})}}{(r^2 +
(z-z^{\prime})^{2})^{3/2}}. \label{march3}
\end{equation}
We can define $s=|z-z^{\prime}|$ and $w = \sqrt{r^2 + s^2}$. Then,
\begin{equation}
I (s) = 2 \pi \int_{s}^{\infty} dw\, \frac{\exp{(- (\chi_{1} /
\alpha_{0}) w)} - \exp{(- (\chi_{2}/ \alpha_{0}) w )}}{w^2}.
\label{march4}
\end{equation}
Defining the quantities
\begin{equation}
\varphi_{0} = 2 \pi S_{0} = \frac{1}{2} \frac{p_{q}}{p + p_{q}} \, \xi \left(
\frac{\nu_{*}}{\nu_{i}} \right) \frac{1}{\ln{(\chi_{max} /
\chi_{min})}}, \label{7abril1}
\end{equation}
and
\begin{equation}
k (s) = \frac{I(s)}{2 \pi} , \label{7abril2}
\end{equation}
we can write the dimensionless photoionization term in the planar
case as
\begin{equation}
S (z)= \varphi_{0} \int dz^{\prime} \, n_e (z^{\prime},t) |{\bf E}
(z^{\prime},t)| e^{-1/|{\bf E} (z^{\prime},t)|} \, k
(z-z^{\prime}) , \label{7abril01}
\end{equation}
where
\begin{equation}
k (s) = \int_{s/ \alpha_{0}}^{\infty} dx\, \frac{\exp{(-\chi_{1}
x)} - \exp{(-\chi_{2} x )}}{\alpha_0 x^2}. \label{7abril02}
\end{equation}
The function $k(s)$ cannot be computed explicitly in terms of
elementary functions, but its asymptotic behaviour can be
calculated. For $s \rightarrow \infty$, we have
\begin{equation}
k (s) \simeq \frac{e^{-(\chi_{1} / \alpha_{0})
s}}{(\chi_{1}/\alpha_{0}) s^2} - \frac{e^{- (\chi_{2}/
\alpha_{0}) s}}{(\chi_{2}/\alpha_{0}) s^2}, \label{7abril02a}
\end{equation}
and for $s \rightarrow 0$, it is
\begin{equation}
k (s) \simeq \frac{\chi_{1} - \chi_{2}}{\alpha_0} \ln{s} + \mbox{const}.
\label{7abril02b}
\end{equation}
In the numerical computations, we will approximate the function
$k(s)$ by functions with the same behaviour at infinity and zero as
the ones shown in equations (\ref{7abril02a}) and
(\ref{7abril02b}). The simulations show that the result is
insensitive to the details of these approximations and they only
depend on the behaviour at zero and infinity. In fact, we will use
a kernel such that it is equal to (\ref{7abril02b}) for $s < 1$
and it is equal to (\ref{7abril02a}) for $s > 1$. The constant in
equation (\ref{7abril02b}) will be chosen in such way that $k(s)$
is continuous at $s=1$.

%\begin{equation}
%\mbox{cte} = \chi_{1} \frac{e^{-(\chi_{1} /
%\alpha_{0})}}{(\chi_{1}/\alpha_{0})^2} - \chi_{2} \frac{e^{-
%(\chi_{2}/ \alpha_{0})}}{(\chi_{2}\alpha_{0})^2}.
%\label{7abril02c}
%\end{equation}

Following \cite{liu} and \cite{Kuli}, we will take for the
simulations $\xi (\nu_{*}/\nu_{i}) = 0.1$, $p_q =
30\,\mbox{Torr}$,
$\chi_{1}=0.035\,\mbox{Torr}^{-1}\mbox{cm}^{-1}\,p_{O_2}$,
$\chi_{2}=2\,\mbox{Torr}^{-1}\mbox{cm}^{-1}\,p_{O_2}$. We will
assume the partial pressure of the oxygen in air is given by
$p_{O_2} = \gamma p$, where $p$ is the total pressure and $\gamma$
a pure number between zero and one. For the inverse ionization
length $\alpha_0$, we will take the value for nitrogen, that
depends on pressure \cite{aftprl} as
$\alpha_0=5.8\,\mbox{Torr}^{-1}\mbox{cm}^{-1}\,p$. For the
diffusion coefficient \cite{Vit}, we take $D_e = 0.1\,{\mbox
m}^2/\mbox{s}$.

Using these values it turns out,
\begin{equation}
\varphi_{0} = 0.37\frac{1}{30+p},
\label{phi0sim}
\end{equation}
with $p$ expressed in Torr, and
\begin{eqnarray}
k(s)&=&\left\{\begin{array}{ll}
  \frac{\exp(-0.006\, \gamma s)}{(0.006\, \gamma) s^2} -\frac{\exp(-0.34\,
  \gamma s)}{(0.34\, \gamma) s^2} ,& \; s > 1,\\
 -0.34\,\gamma \ln{s} + \frac{\exp(-0.006\, \gamma)}{(0.006\, \gamma)}
 -\frac{\exp(-0.34\, \gamma)}{(0.34\, \gamma)},& \; s \leq 1,
\end{array}\right.
 \label{kssim}
\end{eqnarray}

\section{Photoionization without diffusion: acceleration of
negative fronts}
We consider the case in which a divergence-free
electric field ${\bf E}_{0} = -E_{0} {\bf u}_{z}$ is set along the
$z$-axis, so that electrons move towards the positive $z$-axis.
Then we take the electric field as ${\bf E} = - E {\bf u}_{z}$,
$E$ being its modulus. so that, in the case in which the diffusion
coefficient is $D=0$, the model can be written as
\begin{eqnarray}
\frac{\partial n_{e}}{\partial t} &=& - \frac{\partial}{\partial
z} \left( n_{e} E \right) + n_e E e^{-1/E} + S ,
\label{7abril3} \\
\frac{\partial n_{p}}{\partial t} &=& n_e E e^{-1/E} + S ,
\label{7abril4} \\
n_{p} - n_{e} &=& - \frac{\partial E}{\partial z} .
\label{7abril5}
\end{eqnarray}
Now, following the approach presented in
\cite{precorto,japm}, we introduce the shielding factor
$u(z,t)$ as
\begin{equation}
u (z,t) = e^{- \int_{0}^{t} n_{e} (z,t^{\prime} ) dt^{\prime} },
\label{7abril6}
\end{equation}
in terms of which,
\begin{eqnarray}
n_{e} &=& - \frac{1}{u} \frac{\partial u}{\partial t},
\label{7abril7} \\
n_{p} &=& - \frac{1}{u} \frac{\partial u}{\partial t} -
\frac{\partial E_{0} u}{\partial z}  ,
\label{7abril8} \\
E &=& E_{0} u, \label{7abril9}
\end{eqnarray}
and hence
\begin{equation}
S (z)= \varphi_{0} \int dz^{\prime} \, n_e (z^{\prime}) E_{0}
(z^{\prime}) u(z^{\prime}) e^{-1/E_{0} (z^{\prime}) u(z^{\prime})}
\, k (z-z^{\prime})  = -\varphi_{0} \frac{\partial}{\partial t}\int
dz^{\prime} \, G(u(z^{\prime})) \, k(z-z^{\prime}),
\label{7abril10}
\end{equation}
where
\begin{equation}
G (u)= - \int_{u}^{1} du_{1} \, E_{0} e^{-1/E_{0} u_{1}} .
\label{7abril11}
\end{equation}
In order to deduce an equation for the shielding factor $u$, we
follow the steps of \cite{precorto,japm} and obtain a Burgers
equation with non-local source
\begin{eqnarray}
\frac{\partial u}{\partial t} &+& E_{0} u \frac{\partial
u}{\partial z}= - u n_{p0} + u G(u) + \varphi_{0} u \int
G(u(z^{\prime})) \, k (z-z^{\prime}),
\label{7abril11b} \\
u(z,0) &=& 1, \label{7abril12}
\end{eqnarray}
where $n_{p0}$ is the initial positive ion density. Our method of
solution of the above system is by integration along
characteristics; i. e. we solve the following system of ODE's
\begin{eqnarray}
\frac{dz}{dt} &=&  E_{0} u , \label{7abril13} \\
\frac{d u}{d t} &=& - n_{p0} u + u G (u) + \varphi_{0} u \int
dz^{\prime} \, G(u(z^{\prime})) \, k(z-z^{\prime}).
\label{7abril14}
\end{eqnarray}
We use this formulation in terms of characteristics in order
to give a numerical algorithm and study the effect of
photoionization on the propagation of negative planar fronts. We
discretize the spatial variable $z$ into $N$ segments
separated by the points $z_{0}$, $z_{1}$, $\ldots$ $z_{N}$ and
follow the evolution in time of each of them by solving
(\ref{7abril13}) and (\ref{7abril14}). The integral term in
(\ref{7abril14}) is discretized in the following form
\begin{equation}
\int dz^{\prime} \, G(u(z^{\prime}))\, k (z-z^{\prime}) \simeq
\sum_{j=0}^{N-1} G(u(z_{j}(t))) \, k (z_{i} (t) -z_{j} (t)) \,
\left( z_{j+1} (t) - z_{j} (t) \right). \label{7abril15}
\end{equation}

\begin{figure}
 \centering
\includegraphics[width=0.5\textwidth]{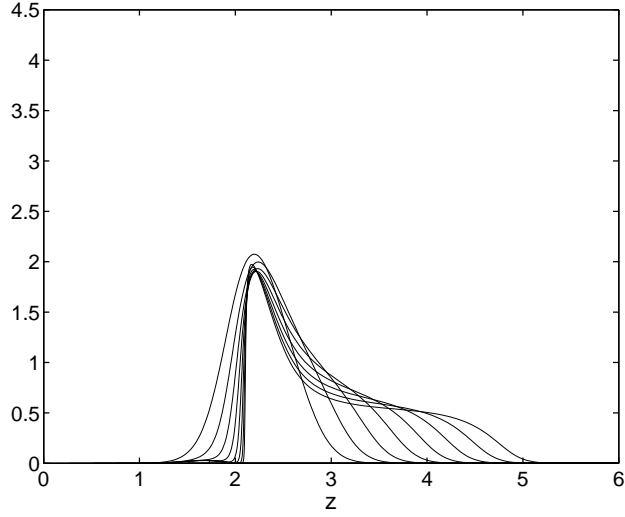}
\caption{Electron density $n_e$ profiles without photoionization.
The electrons move to the right following the polarity of the
electric field. A negative planar front is developed.}
 \label{nophoto}
\end{figure}

\begin{figure}
 \centering
\includegraphics[width=0.5\textwidth]{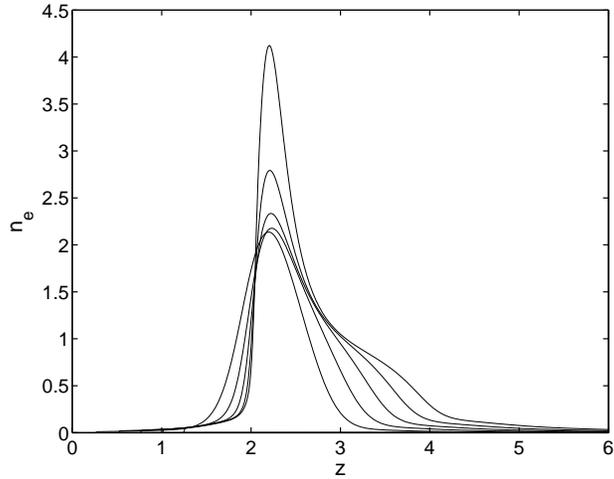}
\caption{Electron density $n_e$ profiles with photoionization, at
normal pressure $p=750\,\mbox{Torr}$ and $\gamma = 0.21$. A
negative front is moving towards the anode at the right and
electrons start getting accumulated at the zero electric field
plasma zone.}
 \label{wphoto}
\end{figure}

In our first numerical experiment, we choose as initial data a
Gaussian distribution of charge. We take $E_0=1$ and the pressure
$p=750\,\mbox{Torr}$. In Fig.\ref{nophoto} we can see the
evolution of the initial negative charge distribution when the
photoionization term is neglected. It can be seen that electrons
move in the direction of increasing $z$ where the anode is
situated. A negative front is developed at the right of the
initial distribution \cite{japm}. The electrons at the left side
of the initial distribution move also following the electric
field, until they reach the main body of the plasma where the
electric field is screened. Then they stop there (around $z=2$ in
Fig.\ref{nophoto}). When the photoionization term is included, the
profiles change. In Fig.\ref{wphoto} the same numerical experiment
is carried out, with the inverse of photoionization range $\gamma
= 0.21$, which corresponds to the normal conditions of air in the
atmosphere.

\begin{figure}
 \centering
\includegraphics[width=0.5\textwidth]{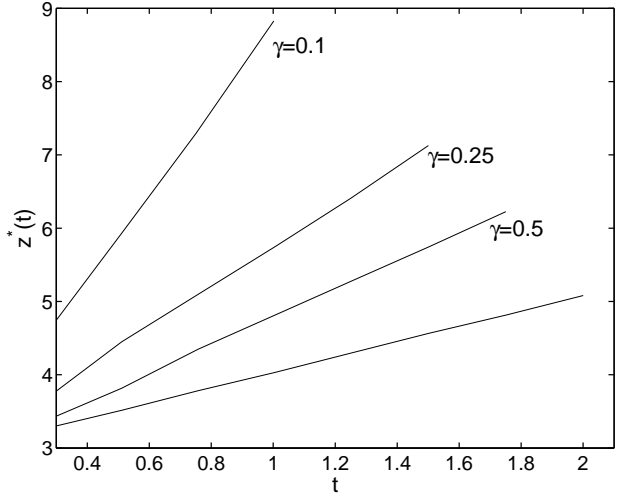}
\caption{The evolution of a point $z^{*}$ of the negative front at which the
electron density has the value $n_e = 0.1$. When photoionization range $1/\gamma$
is increased, the front moves faster. The line without label belongs to the case
where photoionization is neglected.}
 \label{nvelocity}
\end{figure}

We can track the motion of the negative front by looking at the
time evolution of the point $z^{*} (t)$ at which the electron
density has a given value. In Fig.\ref{nvelocity}, we compare the graphs of
$z^{*} (t)$ with and without photoionization for a level of $n_{e}
= 0.1$. As we can see, the effect of photoionization is an
acceleration of the negative front which reaches a higher though
still constant velocity. This fact holds, after our observations,
when one considers kernels $k(s)$ which decay
exponentially fast at infinity.

%If one relaxes this assumption, by allowing a power law decrease of the kernel at infinity,
%then the observation is that the velocity increases indefinitely.

Finally, it is interesting to observe the behaviour of the density
$n_{e}$ in the direction opposed to the propagation of the negative
front (the left part of the initial distribution). This will be called
from now on ``the positive front''. We can observe in Fig.\ref{wphoto}
an effect consisting in the accumulation of electrons in a small
region of space in the positive front. This fact is easy to understand
by considering the production of electrons away from the positive
front which are drifted towards the positive front following the
electric field. In the positive front, electrons and positive ions are
balanced and hence the net electric field cancels. Therefore electrons
cannot proceed any further beyond the positive front and they
accumulate there. This is an effect purely associated to
photoionization which cannot be explained by invoking any different
effect. Unless there is some mechanism allowing the electrons to
spread out once they accumulate at the positive front, their density
will grow indefinitely and eventually will blow up. We will see in
next section that this mechanism is diffusion and the net effect of
photoionization and diffusion is the appearance of travelling waves
moving towards the cathode, i.e. positive ionization fronts.

\section{Photoionization with diffusion: positive ionization fronts}
In this section we study in one space dimension the combined
effect of photoionization and diffusion on the propagation of
positive fronts. The system of equations we study is therefore
\begin{eqnarray}
\frac{\partial n_{e}}{\partial t} &=& - \frac{\partial}{\partial
z} \left( n_{e} E - D \frac{\partial n_{e}}{\partial z} \right) +
n_e E e^{-1/E} + S ,
\label{7abril30} \\
\frac{\partial n_{p}}{\partial t} &=& n_e E e^{-1/E} + S ,
\label{7abril31} \\
n_{p} - n_{e} &=& - \frac{\partial E}{\partial z}
,\label{7abril32}
\end{eqnarray}
where $S$ is the photoionization source term and is written as in
equation (\ref{7abril10}).

The main difference in our approach to this problem with respect
to the problem without diffusion is that now an integration along
characteristics does not lead to simplifications due to the
presence of the second derivatives associated with diffusion.
Instead we will use the method of finite differences.

\begin{figure}
 \centering
\includegraphics[width=0.5\textwidth]{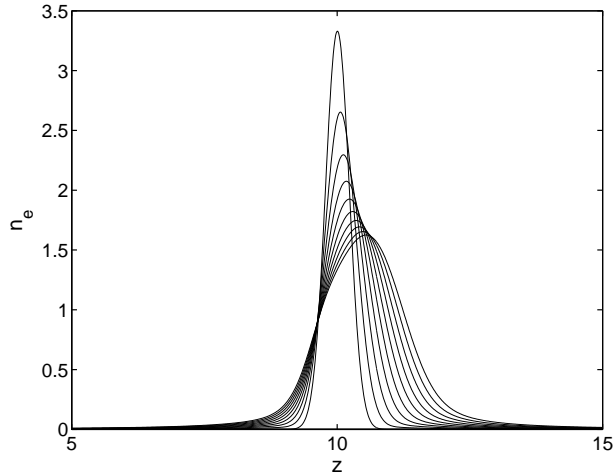}
\caption{Electron density $n_e$ profiles, at normal pressure
$p=750\,\mbox{Torr}$, photoionization parameter $\gamma = 0.25$
and diffusion $D=0.57$ in dimensionless units. A negative front is
moving towards the anode at the right and a positive fronts
towards the cathode at the left.}
 \label{dphoto}
\end{figure}

\begin{figure}
 \centering
\includegraphics[width=0.5\textwidth]{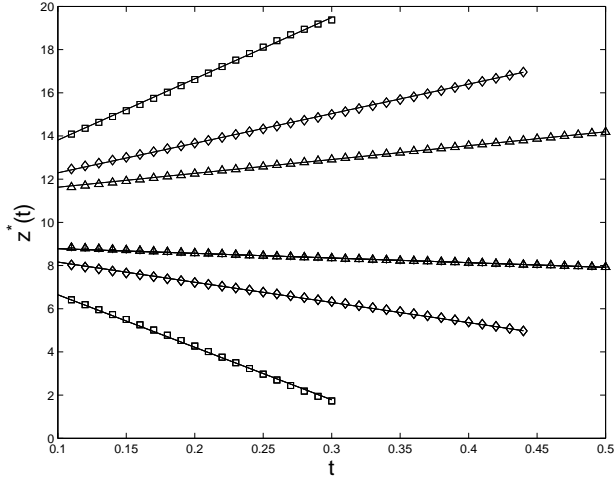}
\caption{The evolution of points $z^{*}$ of the negative and positive
fronts at which the electron density has the value $n_e = 0.02$. The increasing
values are for the negative front and the decreasing ones are for the positive.
When photoionization range $1/\gamma$ is increased, the fronts move faster.
Triangles $\triangle$ are for $\gamma=0.9$, diamonds $\Diamond$ for $\gamma=0.25$
and squares $\Box$ for $\gamma=0.1$.}
 \label{dvelocity}
\end{figure}

In Fig.\ref{dphoto}, we represent the profiles for $n_{e}$ with
$D=0.57$, $p=750\,\mbox{Torr}$ and $\gamma=0.25$. We have used an
initial charge distribution which has a maximum at $z=10$. When it
evolves, it can be observed a negative planar front developing.
The propagation of the negative front is almost identical with or
without diffusion when photoionization is present. However there
is now a positive front moving towards the cathode. The positive
front moves with a constant velocity which is smaller than the
velocity of the negative front. In Fig.\ref{dvelocity} we have
plotted the position $z^*$ of a point of the negative front and of
the positive front which has the particular value of the electron
density $n_e = 0.02$. The parameters are the same as in
Fig.\ref{dphoto}, but for three different values of $\gamma$. For
the parameter values chosen above, we have computed the ratio
between the velocities of positive and negative fronts:
$c_{pos}/c_{neg} = 0.34$ for $\gamma = 0.9$, $c_{pos}/c_{neg} =
0.68$ for $\gamma = 0.25$ and $c_{pos}/c_{neg} = 0.86$ for $\gamma
= 0.1$. The ratio grows when the photoionization range $1/\gamma$
increases and the velocities for negative and positive fronts tend
to increase and get closer to each other.

The propagation of positive fronts as travelling waves results from
the combined action of photoionization and diffusion. This is in
contrast with the propagation mechanism for negative fronts, which are
also travelling waves but they result from a combination of impact
ionization and convection by the electric field. In the latter case,
diffusion and photoionization only affect the negative fronts by
changing their velocity and their shape. All this conclusions are
rather insensitive to the detailed form of the kernel $k(s)$
(see formula (\ref{7abril2})) provided it decays exponentially fast at
infinity, and hence our conclusions hold with a high degree of generality.

\section{Conclusions}
In this paper we have studied the effect of photoionization in
streamer discharges. We have deduced a minimal model including
photoionization and studied with this model the propagation of
both positive and negative fronts in the planar case. We have
found the appearance of travelling waves which accelerate when the
photoionization range increases. For negative fronts we have
studied the effect of photoionization both when electronic
diffusion is neglected and included. For positive fronts,
electronic diffusion has to be taken into account and we have
shown how photoionization plays the crucial role pointed by
Raether on increasing the velocity of propagation. The control
parameter is the photoionization range, i.e. the typical distance
at which photons are able to ionize the media. Physically in air,
this parameter depends on the amount of oxygen and nitrogen
present. It is interesting to point out that for real discharges
in the atmosphere, this parameter varies with the altitude.


\begin{thebibliography}{99}
\bibitem{Raether} H. Raether, Die Entwicklung der
Elektronenlawine in den Funkenkanal, {\it Z. Phys.} {\bf 112}, 464--489 (1939).
\bibitem{Loeb} L. B. Loeb, The problem of the mechanism of static spark discharge, {\it Rev.
Mod. Phys.} {\bf 8}, 267--293 (1936).
\bibitem{cp} M. Array\'as and J. L. Trueba,
Investigations of Pre-Breakdown Phenomena: Streamer Discharges,
{\it Cont. Phys.} {\bf 46}, 265--276 (2005).
\bibitem{Rai} Y. P. Raizer, {\it Gas Discharge Physics} (Springer, Berlin
1991).
\bibitem{LoMe} L. B. Loeb and J. M. Meek, {\it The mechanism of
the electric spark}, Clarendon Press, Oxford, 1941.
\bibitem{Pasko} V. P. Pasko, M. A. Stanley, J. D. Mathews, U. S. Inan, and T. G. Wood,
Electrical discharge from a thundercloud top to the lower ionosphere, {\it Nature} \textbf{416}, 152--154 (2002).
\bibitem{ME} M. Array\'as, U. Ebert, and W. Hundsdorfer,
Spontaneous branching of anode-directed streamers between planar
electrodes, {\it Phys. Rev. Lett.} {\bf 88}, 174502 (2002).
\bibitem{aftprl} M. Array\'as, M. A. Fontelos, J. L. Trueba,
Mechanism of branching in negative ionization fronts, {\it Phys.
Rev. Lett.} {\bf 95}, 165001 (2005).
\bibitem{Dhali} S. K. Dhali and A. P. Pal, Numerical simulation
of streamers in SF$_6$, {\it J. Appl. Phys.} {\bf 63}, 1355--1362
(1988).
\bibitem{liu} N. Liu and V. P. Pasko, Effects of photoionization
on propagation and branching of positive and negative streamers in
sprites, {\it J. Geophys. Res.} {\bf 109}, A04301 (2004).
\bibitem{naidis} G. V. Naidis, On photoionization produced by discharges on air,
{\it Plasma Surces Sci. Technol.} {\bf 15}, 253--255 (2006).
\bibitem{Zhe} M. B. Zhelezniak, A. Kh. Mnatsakanian, S. V. Sizykh, Photoionization of nitrogen and oxygen
mixtures by radiation from a gas discharge, {\it High
Temperature}, {\bf 20}, 357--362 (1982).
\bibitem{Vit} P. A. Vitello, B. M. Penetrante, and J. N. Bardsley,
Simulation of negative-streamer dynamics in nitrogen, {\it Phys.
Rev. E} {\bf 49}, 5574--5598 (1994).
\bibitem{Ute} U. Ebert, W. van Saarloos, and C. Caroli, Streamer
propagation as a pattern formation problem: Planar fronts, {\it
Phys. Rev. Lett.} {\bf 77}, 4178--4181 (1996), and ibid.,
Propagation and structure of planar streamer fronts, {\it Phys.
Rev. E} {\bf 55}, 1530--1549 (1997).
\bibitem{AJP} M. Array\'as, On negative streamers: A deterministic
approach, {\it Am. J. Phys.}
{\bf 72}(10), 1283--1289 (2004).
\bibitem{precorto} M. Array\'as, M. A. Fontelos, J. L. Trueba, Ionization
fronts in negative corona discharges, {\it Phys. Rev. E} {\bf 71}, 037401 (2005).
\bibitem{japm} M. Array\'as, M. A. Fontelos, J. L. Trueba, Power laws and
self-similar behaviour in negative ionization fronts, {\it J. Phys. A: Math. Gen.} {\bf 39}, 1--18 (2006).
\bibitem{Kuli} A. A. Kulikovsky, The role of photoionization in positive streamer dynamics,
{\it J. Phys. D: Appl. Phys.} {\bf 30}, 1514--1524 (2000).
\end{thebibliography}
\end{document}